\pdfoutput=1

\documentclass[10pt,letterpaper,twocolumn]{article} 

\usepackage{ol2}
\usepackage[draft]{hyperref}
\usepackage{amsmath}

\begin{document}

\twocolumn[ 
\title{Photonic crystal nanofiber using an external grating}
\author{M. Sadgrove, R. Yalla, K.P. Nayak and K. Hakuta}
\address{
Center for Photonic Innovations, The University of Electro-Communications, 1-5-1 Chofugaoka, Chofu, Japan
}

\begin{abstract}
We implement a photonic crystal nanofiber device by reversibly combining an optical nanofiber and a nanofabricated grating.
Using the finite-difference time-domain method, we design the system for minimal optical loss while tailoring the resonant wavelength and 
bandwidth of the device. Experimentally we demonstrate that the combined system shows a strong photonic stop-band 
in good agreement with numerical predictions. The resulting device may be used to 
realize strong light-matter coupling near to the nanofiber surface.
\end{abstract}

]
The realization of strong photonic confinement in micro and nano-scale dielectric structures~\cite{Microcavity} and the 
resultant enhancement of light-matter coupling which can be achieved near the surface of such structures has 
led to the current great interest in nanophotonics~\cite{Nanophotonics}.
Within this field, photonic crystal (PhC) devices~\cite{PhCBook} in particular have shown to be a 
promising technology, with potential applications ranging from guided single photon emission~\cite{Hughes}  
and atom trapping~\cite{Kimble4,Lukin1} to biosensing~\cite{nanolett}.

Another approach which naturally achieves the high mode confinement and 
strong light matter coupling associated with PhCs is that of sub-wavelength tapered optical fibers (nanofibers).
High optical density and high coupling between quantum emitters and the guided-modes of a nanofiber
have recently been demonstrated in such systems~\cite{Kali1, Rauschenbeutel1, Kimble3, Chandra}.
This light-matter coupling can be further enhanced by implementing a photonic crystal 
structure on the nanofiber~\cite{LeKienHakuta2}. 
Experimental realizations of such \emph{photonic crystal nanofibers} have been achieved
using ion beam etching~\cite{KaliFIB} and femtosecond laser ablation techniques~\cite{KaliPCN} to 
fabricate periodic structures on the nanofiber surface.
\begin{figure}[htb]
\centerline{\includegraphics[width=0.95\linewidth]{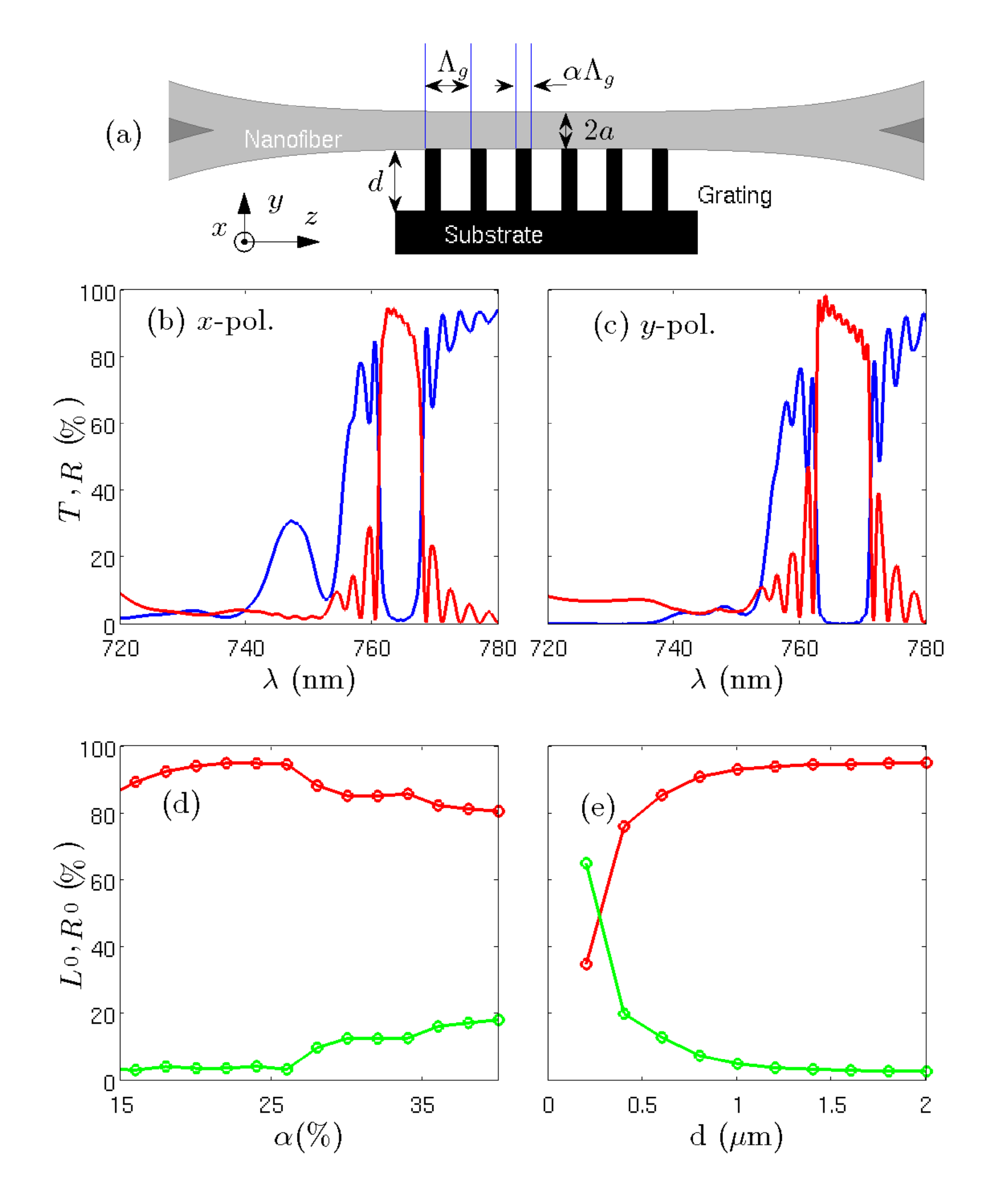}}
\caption{\label{fig:design} (a) Experimental concept. An external grating is combined with an optical nanofiber forming a device with PhC characteristics. 
(b) and (c) show FDTD simulation results for the transmission $T$ (blue line) and reflection $R$ (red line) 
spectra for $x$ and $y$ polarizations respectively with grating parameters as defined in the main text. (d) and (e) show FDTD simulation results for the maximum reflectivity $R_0$ (red line) and optical loss $L_0$ (green line) as a function of the grating duty cycle $\alpha$ and the grating slat depth $d$ respectively. In both cases, the input
mode was $y$-polarized.  }
\end{figure}

In this Letter, we present a new method of implementing photonic crystal properties in an optical nanofiber by 
combining the nanofiber with an external grating as illustrated in Fig.~\ref{fig:design}(a). Our method
takes advantage of modern nanofabrication methods, which allow flexible design of planar nanostructures,
in combination with nanofiber technology which produces strong transverse mode confinement and allows 
strong light-matter coupling and high optical densities for emitters near the nanofiber surface. 
The grating, which is fabricated on a silica substrate, is brought into contact with a suspended nanofiber. 
In the region where the nanofiber and the grating make contact, the effective refractive index experienced by
the nanofiber guided mode is strongly modulated by the grating giving rise to a photonic stop-band. 
The stop-band along with the strong transverse mode confinement due to the nanofiber
mean that the combined device may be considered to be a fiber-coupled PhC.  
Our device represents a new paradigm in the implementation of PhCs where 
two (or more) independent 
nano-scale optical elements may be reversibly combined to form an effective PhC structure.
Combined with a quantum dot deposited on the nanofiber surface~\cite{Chandra} or following recent proposals 
for trapping single atoms in PhCs~\cite{Lukin1,Kimble4},
we believe that our technique can provide a powerful and flexible tool for the realization of strong light-matter 
coupling with applications ranging from quantum and nano optics to biosensing.  

As shown in Fig.~\ref{fig:design}(a) the external grating is defined by the following parameters: grating period $\Lambda_g$,
slat width $\alpha\Lambda_g$, where $\alpha$ is the grating duty cycle, slat depth $d$ and the number of slats $N$.
Before manufacturing the grating, we simulated the system using the finite-difference time-domain 
(FDTD) method (Lumerical Solutions Inc.) to evaluate the optical properties of the device.
In particular, in order to keep the optical loss to a minimum, it was necessary to carefully evaluate different 
duty cycles $\alpha$ and slat depths $d$ for the grating using simulations.
Although a larger duty cycle produces stronger index modulation, and thus stronger reflectivity in the stop-band, it also allows more coupling to the silica substrate
and thus greater optical loss. Likewise, a shallower grating is easier to manufacture, but if the grating is too shallow, the evanescent field extending from the nanofiber can couple to 
the silica substrate leading to large losses. 
 Figures~\ref{fig:design}(b) and (c) show reflection and transmission spectra from FDTD simulations for
 a nanofiber diameter of $2a$ = 625 nm  and an external grating with period $\Lambda_g = 305$ nm, 
duty cycle $\alpha=25\%$, grating depth $d=2\;\mu$m and $N=200$ slats. 
 
 Two orthogonal polarizations $x$ and $y$ were tested, with the $x$ and $y$ directions as defined in Fig.~\ref{fig:design}(a).
 For both polarizations, a strong photonic stop-band is seen in the transmission spectrum near $765$ nm
 accompanied by a peak in the reflection spectrum. The device is seen to be lossy 
 below a wavelength of $\sim760$ nm.
Denoting by $\lambda_0$ the wavelength at which the maximum reflection occurs (i.e. the \emph{working wavelength} of the device),
 we can define $R_0$ and $T_0$, the reflectivity and transmission at $\lambda_0$ respectively. The optical loss at the working
wavelength is then $L_0 = 1-R_0-T_0$.
 For the $x$ polarization, we have $\lambda_0=762$ nm and $R_0=95\%$ and the reflection spectrum
 full-width at half-maximum (FWHM) is $\sigma_R = 7$ nm. $T_0$ and $L_0$ are found to be $1\%$ and $4\%$ respectively.
 For the $y$ polarization we find $R_0 = 98\%$ and $\sigma_R = 9$ nm, with $T_0 < 0.1\%$ and $L_0$ = $2\%$.
We note that the polarization dependence of the device is not very pronounced, with similar stop-bandwidths and peak reflectivities
seen for both polarizations. 
 
To optimize the optical properties of the device, we performed parameter sweeps over $\alpha$ and $d$ in our FDTD simulations.
 Figures~\ref{fig:design}(d) and (e) show the results for $L_0$ and $R_0$ as functions of $\alpha$ and $d$ respectively for grating and fiber parameters as 
given above. For the duty cycle sweep, we set $d=2\;\mu$m, and for the depth sweep we set $\alpha=25\%$ and the input mode was $y$-polarized in both cases.
As can be seen in Fig.~\ref{fig:design}(d), above $\alpha=25\%$, a relatively sharp increase in loss is observed, and indeed we found that the loss behaviour
at higher $\alpha$ displayed more complex characteristics as a function of the input wavelength $\lambda$. One possible reason for the loss behaviour
above $\alpha = 25\%$ is the waveguide properties of the slats themselves. Clearly, as the slat width $\alpha\Lambda_g$ increases, the number of modes which can couple to the slats 
becomes larger leading to more complicated loss behavior.   
On the other hand, in Fig.~\ref{fig:design}(e) it may be seen that the loss decreases smoothly as a function of $d$, falling to $L_0=2\%$ at $d=2\;\mu$m.
\begin{figure}[htb]
\centerline{\includegraphics[width=1.1\linewidth]{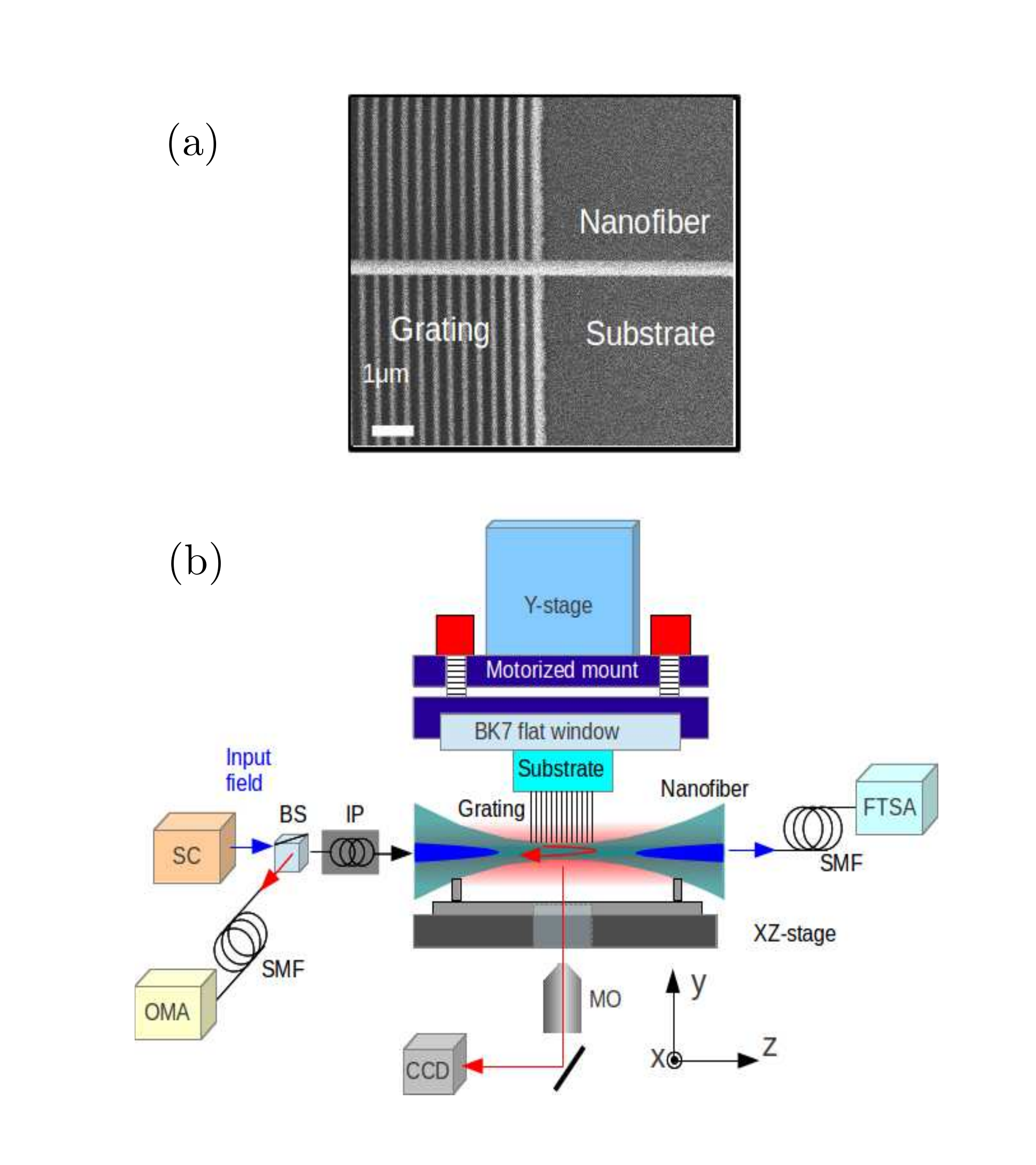}}
\caption{\label{fig:exp} (a) SEM image of the device clearly showing the grating and substrate regions along with the nanofiber. 
(b) Schematic diagram of the experimental setup. 
Light from a super-continuum source (SC) is coupled 
into the nanofiber which is fixed on an XZ-stage mounted on an inverted microscope. The output transmission and reflection spectra are measured by
a Fourier transform spectrum analyzer (FTSA) and an optical multichannel analyzer (OMA) respectively. A full explanation of all the symbols is provided in the text.} 
\end{figure}
After finding the optimum parameters for $\alpha$ and $d$, a similar optimization procedure was used to choose $N$, with $N=200$ found to be optimal.
These results demonstrate that an external grating can be used to induce a photonic stop-band in a nanofiber
with high reflectivity ($R_0>95\%$) and with optical loss as low as $2\%$ for the parameters considered here.

For the experiments, we used a 200 slat, square grating pattern with a nominal period of 320 nm and a duty cycle
of 25$\%$ fabricated on a 10$\;$mm$\times$25$\;$mm silica substrate of thickness 2$\;$mm (fabricated by Nanoeff Consultant ltd.). 
The grating pattern extended $d=2\;\mu$m from the surface of the silica substrate. Figure~\ref{fig:exp}(a) shows 
a scanning electron microscope (SEM) image of the grating with a $570$ nm diameter nanofiber mounted on it.

The experimental setup is illustrated in Fig.~\ref{fig:exp}(b). 
The grating substrate was attached to a one inch diameter optically flat BK7 window
using an adhesive. The mounted grating was then inserted into a high resolution, 
motorized mirror mount (Newfocus Picomotor 8884) so that its tilt could be precisely 
controlled. The motorized mount was attached to a vertical translation stage (Y-stage) 
(Kohzu XD2000) with 250$\;$nm resolution. So mounted, the grating 
could be lowered onto the nanofiber sample with high precision. The nanofiber 
itself was mounted on a two-axis translation stage (XZ-stage) (Sigma Koki BIOS 305T)
with a resolution of 50 nm to allow adjustment of the relative centers of the grating and the nanofiber. 
The XZ-stage and the Y-stage were mounted on an inverted microscope (Nikon Eclipse Ti-U).
The whole experiment took place in a clean booth to avoid dust deposition on the grating and nanofiber.

For characterization of the device's optical properties we used an optical setup 
as described below.
The input to the nanofiber was derived from a filtered supercontinuum light source (SC) 
(NKT Photonics SuperK EXR15). The filtered spectrum spanned from 700-1000 nm
and the light was passed through a polarizing beam-splitter to ensure linear polarization.
The polarization angle of the input light was controlled by an inline polarizer (IP) (Thorlabs FPC030).
The transmission and reflection spectra were measured by a Fourier transform spectrum analyzer (FTSA) (Thermo Fisher Scientific Nicolet 8700) with 
a resolution of $0.01\;$nm, and an
optical multichannel analyzer (OMA) (Ocean optics QE65000) with resolution 2$\;$nm respectively.
We also monitored the nanofiber/grating interface using the microscope's 40x objective lens (MO) which 
focused an image of the device to a charge coupled device (CCD) camera. 
\begin{figure}[htb]
\centerline{\includegraphics[width=0.8\linewidth]{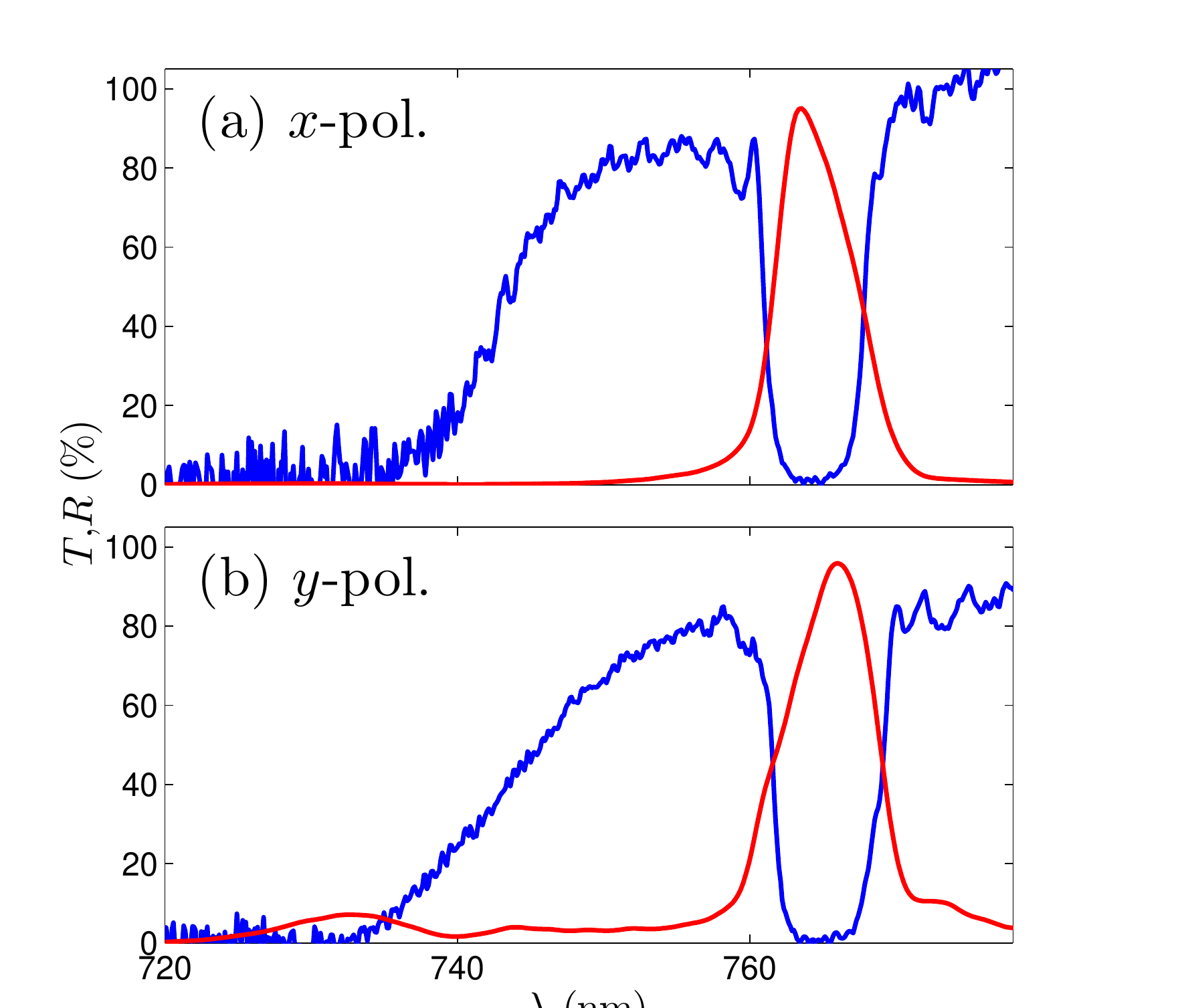}}
\caption{\label{fig:expres}  (a) and (b) show experimentally measured transmission (blue line) and reflection (red line) spectra for the grating placed near the nanofiber center
(nominal nanofiber diameter 570$\;$nm) for $x$ and $y$ polarized light respectively. 
}
\end{figure}

Before bringing the grating into contact with the nanofiber, we measured the tilt of the nanofiber with 
respect to the horizontal using the microscope.
We then adjusted the offset of the precision motorized mount so that the grating tilt matched that of the nanofiber.
Finally, we used the vertical translation stage to bring the grating close to the nanofiber.
Optical contact between the grating and the nanofiber was marked by the appearance of scattered light
at the nanofiber surface (as seen on the CCD camera). From this point, successful mounting of the 
grating on the nanofiber was accomplished by fine adjustments to the Y-stage position while monitoring
the signals of the FTSA and OMA.

Figures~\ref{fig:expres}(a) and (b) show measured transmission and reflection spectra for two orthogonal polarizations
for a nanofiber diameter of 570$\;$nm. We see a strong optical stop-band in the transmission spectrum at a wavelength 
of 765 nm accompanied by a peak in the spectrum of the reflected light occuring at $\lambda_0=764\;$nm and 766 nm for the $x$ and $y$ polarizations respectively.
We measured the bandwidth $\sigma_R$ for the reflection spectra shown in
Figs.~\ref{fig:expres}(a) and (b) and found  $\sigma_R = 6\;$nm for the $x$ polarization and 
$\sigma_R = 7\;$nm for the $y$ polarization. The optical loss is seen to become large below 
$\lambda\sim750$ nm.

In order to calibrate $R_0$, $T_0$ and $L_0$ for the experimental spectra, we replaced the super-continuum light source with 
a Titanium Sapphire laser source (Coherent MBR 110, linewidth $<75\;$kHz) and measured the reflected and transmitted 
power through the system  at $\lambda_0$ and several other wavelengths within and just outside the optical stop-band. Five measurements were made at each wavelength 
both with and without the grating touching the nanofiber.
The calibrated spectrum for the $x$ polarization was found to have a maximum reflection $R_0=(96\pm 1)\%$, 
while for the $y$ polarization we found $R_0=(95\pm4)\%$. For both polarizations, the transmission $T_0$ at the peak reflective wavelength 
was less than $0.1\%$ giving losses of $L_0=(4\pm1)\%$ and $L_0=(5\pm4)\%$ for the $x$ and $y$ polarizations respectively. For all values,
the given errors are the standard deviation over five measurements.

We see good qualitative agreement between experimental results and the spectra calculated using FDTD numerical simulations.
The quantitative values of the maximum reflectivity $R_0$ and $\sigma_R$ are also in fair agreement between simulations and 
the experimental spectra as are the values of $L_0$ and $T_0$. 
Additionally, both simulation and experimental results show the same lossy region for wavelengths below 750$\;$nm.
In order to reproduce both the qualitative loss characteristics and the resonant wavelength and bandwidth 
of the experimental spectra, it was necessary to use a grating period of $\Lambda_g = 305\;$nm and 
a nanofiber diameter of $2a=625\;$nm in the simulations. The difference between these parameters and the 
nominal experimental parameters is not yet fully understood. However, we note that the experimental
values were determined by SEM measurements which have an error of $\pm 10\%$.
The values used in the simulations lie within this error range.
\begin{figure}[htb]
\centerline{\includegraphics[width=0.9\linewidth]{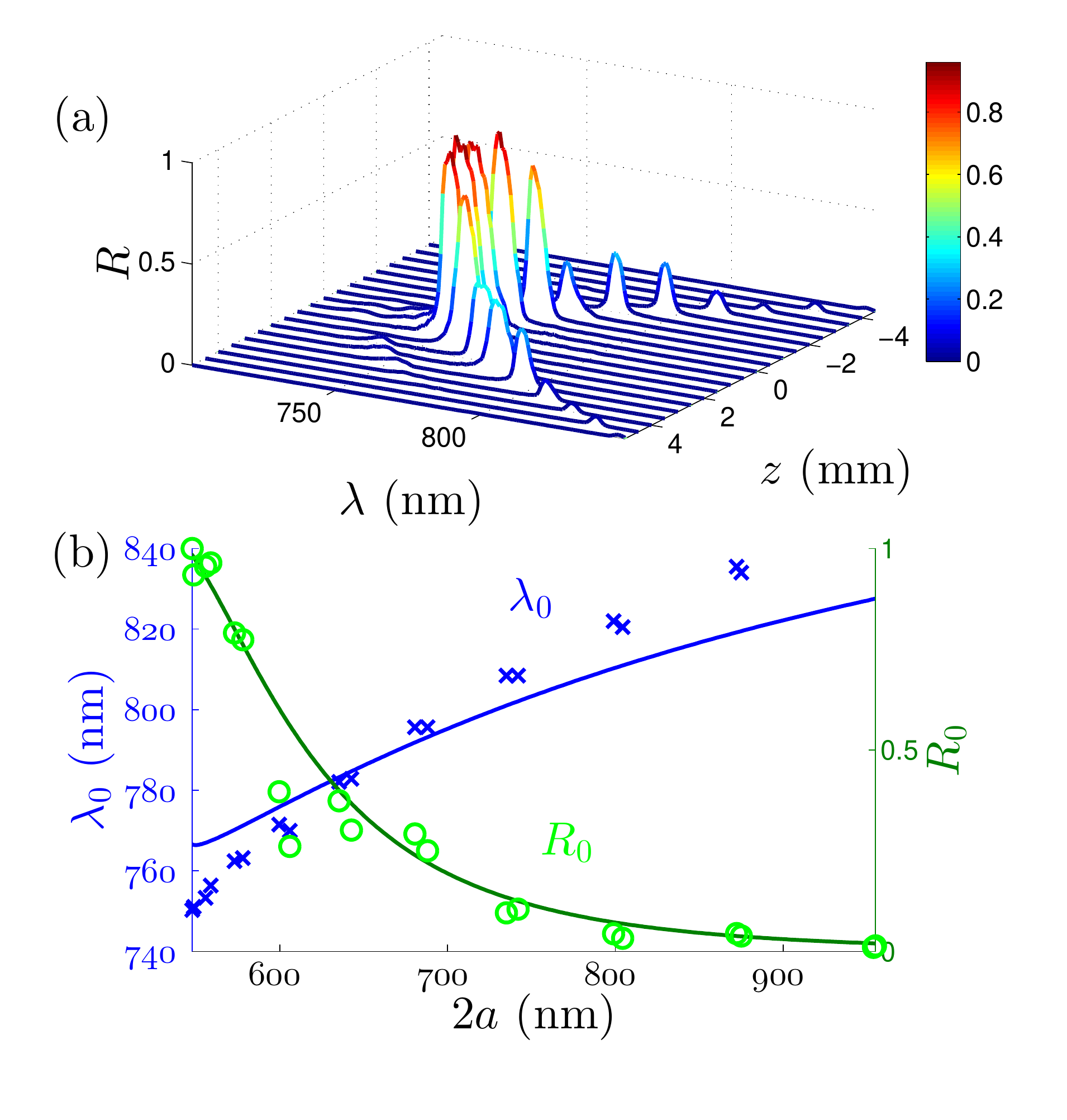}}
\caption{\label{fig:radius} 
In (a), reflection spectra taken at a number of different $z$ values are shown.
(b) shows the peak reflection wavelength (blue crosses) $\lambda_0$ 
and the peak height $R_0$ (green circles) as a function of the nanofiber diameter $2a$. 
The predicitons of coupled mode theory are shown for $\lambda_0$ by a solid blue line and 
for $R_0$ by a solid green line.
}
\end{figure}

We also tested the dependence of the resonant wavelength on the diameter of the nanofiber by placing the grating at various positions
along the nanofiber ($z=0$ corresponds to the thinnest part of the nanofiber where $2a=550$ nm). 
The results are shown in Figs.~\ref{fig:radius} (a) and (b). 
The resonant peak position $\lambda_0$ is seen to shift to the red in proportion to the change in nanofiber diameter.
The falloff of the reflection peak height $R_0$ as $a$ increases (green circles, Fig.~\ref{fig:radius}(b)) 
is due to the reduction in the effective index modulation at thicker diameter.

It is meaningful to evaluate our device using the coupled mode theory (CMT)~\cite{Erdogan}  of fiber Bragg gratings. 
If we model the the index modulation by a sinusoidal variation $n=n_0+\Delta n\sin^2[(\pi/\Lambda_g)z]$, where $n_0$ is the unperturbed refractive index of the core and $\Delta n$ is the modulation amplitude, CMT gives $\lambda_0=\lambda_D(1+\Delta n/2)$, where $\lambda_D=2n_{\rm eff}\Lambda_g$, 
with $n_{\rm eff}$ the effective refractive index seen by the nanofiber guided mode. The peak reflection value is given by $R_0=\tanh^2(2\pi\Delta n N\Lambda_g/\lambda_0)$. We evaluated $\Delta n$ empirically by fitting 
the CMT theory to our experimental results, assuming the same parameters as used in the FDTD simulations. 
$\Delta n$ was found to have a maximum of $1.1\%$ at $2a=550$ nm and dropped to almost 0 at $2a=950\;$nm. Since $n_{\rm eff}$ increases as the core diameter gets larger, $\lambda_0$ is expected to move to the red as shown by the solid blue line in Fig.~\ref{fig:radius}(b),
while $R_0$ decreases as shown by the solid green line in Fig.~\ref{fig:radius}(b). 
It may be seen that the essential experimental features are reproduced, indicating that CMT is a useful tool for the proto-design of the device. 

Our device may also be considered to be a 1D PhC waveguide. It is well known that for a quantum emitter
placed near to such a waveguide, Purcell enhancement of emission into the guided modes of the waveguide is expected~\cite{Hughes}.
We have performed preliminary FDTD simulations which suggest a Purcell factor of order 10 is possible for the device introduced here.
 Corresponding coupling to the guided modes of the nanofiber is then expected to be more than $60\%$.
In addition, engineering a sub-wavelength dimension cavity in the grating pattern is another method to enhance coupling
between a quantum emitter and the nanofiber guided modes.
Such a structure can create a strong cavity mode at the center of the stop-band in a way analogous to 
``defect-mode" cavities used in distributed feedback lasers~\cite{Painter}. We note that for a quantum dot deposited on the nanofiber
as in Ref.~\cite{Chandra}, the Purcell enhancement can be optimized by adjusting the relative position of the grating and the nanofiber
which is an advantage of our method. 

The present work demonstrates that PhC structures with low optical loss can be made by combining two separate
nano-scale optical elements. The combined device is a PhC nanofiber with high transverse mode confinement due to the nanofiber and a stop-band due to the 
external  grating. With suitable parameter choices, it should be possible
to use the device to realize strong light-matter coupling between the nanofiber guided modes and quantum emitters.

We would like to thank Pengfei Zhang for useful discussions. This
work was supported by the Japan Science and Technology Agency (JST) as one of the Strategic
Innovation projects.

\end{document}